\documentclass{iau} 
\usepackage{natbib}
\usepackage{graphicx}

\title[New $\gamma$-ray blazars] 
{A search for new $\gamma$-ray blazars from infrared selected candidates }

\author[Musiimenta Blessing]   
{Musiimenta Blessing$^1$,
 Arsioli Sversut Bruno$^2$, Jurua Edward$^1$ and Mutabazi Tom$^1$}

\affiliation{$^1$Mbarara University of Science and Technology (MUST), Mbarara, Uganda \\ email: {\tt mblessing78@gmail.com} \\[\affilskip]
$^2$Instituto de Fisica Gleb Wataghin IFGW, Unicamp, Brasil }

\pubyear{2019}
\volume{356}  
\jname{Nuclear activity in galaxies across cosmic time}
\editors{M. Povi\'c, P. Marziani, J. Masegosa, H. Netzer,\\ S. H. Negu, \& S. B. Tessema, eds.}
\begin{document}

\maketitle

\begin{abstract}
We present a systematic study of gamma-ray blazar candidates based on a sample of 40 objects taken from the WIBR catalogue. By using a likelihood 
analysis, 26 of the 40 sources showed significant gamma-ray signatures $\geq3\sigma$. Using high-energy 
test statistics (TS) maps, we confirm 8 sources, which are completely new, and show another 15 promising $\gamma$-ray candidates. The results from 
this analysis show that a multi-frequency approach can help to improve the current description of the gamma-ray sky.
\end{abstract}

\firstsection 
\section{Introduction}
\noindent
Blazars are Active Galactic Nuclei (AGN) with relativistic jets pointed towards Earth \citep{urryp1995}. While blazars are rare extragalactic 
sources, they represent the largest fraction of gamma-ray sources \citep{abdo2010a} and contribute to the extragalactic 
gamma-ray background (EGB) \citep{mukherjee1997,abdo2010b,ajello2015}. A complete description of the EGB is still an open 
issue in gamma-ray astronomy. Therefore, searching for new gamma-ray sources from low energy blazar candidates can improve the description of the 
gamma-ray sky \citep{arsioli2017b1}.

\section{Methods}
 
 \noindent
The WIBR catalogue \citep{abrusco2014} contains 7855 sources. Many of them happen to be part of the 2WHSP catalogue \citep{chang2017} and 
5BZcat \citep{massaro2015}. We selected 40 sources from the WIBR catalogue which have no counterparts in the Fermi catalogue, therefore with 
focus on undetected gamma-ray sources. 
\begin{itemize}
 \item [(i)]
 The Spectral Energy Distributions (SEDs) were analysed using the ASDC SEDbuilder tool.
  \item [(ii)]
 The gamma-ray analysis was performed using Fermi science tools.

The significance ($\rm{\sqrt{TS}\,\sigma}$) of the detections was determined using the Test Statistic (TS) parameter defined as:
\begin{equation}
 \rm{TS =-2(\ln L_{\textit{no-source}}-\ln L_{\textit{source}})},
\end{equation}
where $L_{\textit{no-source}}$ is the null hypothesis, and $L_{\textit{source}}$ is 
the likelihood value for a model with the additional candidate source at the
same position \citep{mattox1996}. The gamma-ray spectrum of the sources was assumed to be described by a power law model given by:
 \begin{equation}
\rm{\frac{dN}{dE}=N_o\left(\frac{E}{E_o}\right)^{-\Gamma}},
\end{equation}
where $\rm{E_o}$ is pivot energy, $\rm{N_o}$ is the prefactor (corresponding to the flux density in 
$\rm{ph/cm^2/s/MeV})$ at $\rm{E_o}$  and $\Gamma$ is the photon spectral index for a given energy range.

\item [(iii)]
TS maps were obtained by performing an unbinned likelihood analysis using Fermi science tools.

\end{itemize}

\section{Results and discussion}
 
\subsection{Significant detections}

In 3FGL and 4FGL catalogues, sources were detected at 4$\sigma$ significance \citep[][respectively]{acero2015,4fgl2020}. However, in this analysis, 
sources with $\sqrt{TS}\sim{3}\sigma$ are considered to be $\gamma$--ray detections with low significance following the discussion from 
\cite{arsioli2017b1,arsioli2018b2}. All cases with TS between 9 and 25 are considered to be a relevant excess signature, given that those 
cases actually 
have a multi-frequency counterpart (radio to X-rays) as expected from blazars.

Out of the 26 sources with a significant gamma-ray signature, 10 were detected with TS$\,>\,$25 and 16 were detected with 
9$\,<\,$TS$\,<\,$25. All these sources are entirely new detections out of previous Fermi catalogues.

 \subsection{New WIBR gamma-ray detections}
 
Different TS maps were built with different photon energy cuts i.e., 950 MeV, 1 GeV, 2 GeV, 2.5 GeV, and 3 GeV. The TS maps were built taking into 
consideration the computational time and also to obtain the TS peak with the best resolution to better solve the gamma-ray signature as a point-like 
source. All these sources are at high Galactic latitudes ($|b|>10^\circ$), avoiding the Galactic diffuse emission and also preventing spurious 
detections.

Out of the 26 new WIBR $\gamma$-ray detections, we built TS maps for 23 sources (both high and low significance detections). Three 
sources were not fully analysed due to very high photon counts that increased the computational heaviness. This is because those cases seemed
more relevant at the lowest energy channel from Fermi-LAT ($\rm{E\,<\,900\,MeV}$) and building TS maps in this energy range becomes computationally 
far 
too heavy, especially if necessary to integrate over many years of observations. Out of the 23 TS maps built, 8 show a point-like source 
at the positions of the target sources and 15 sources could not be confirmed with high energy TS maps, but might be relevant at lower energies and 
therefore are flagged as promising $\gamma$-ray candidates. Figure \ref{fig3} (left panel) shows 
a TS map of one of the new gamma-ray sources.

\subsection{Promising gamma-ray candidates}
\begin{figure}[ht]{}
 \begin{center}
 \advance\leftskip -3cm
 \advance\rightskip -3cm
  \includegraphics[width=2.4in]{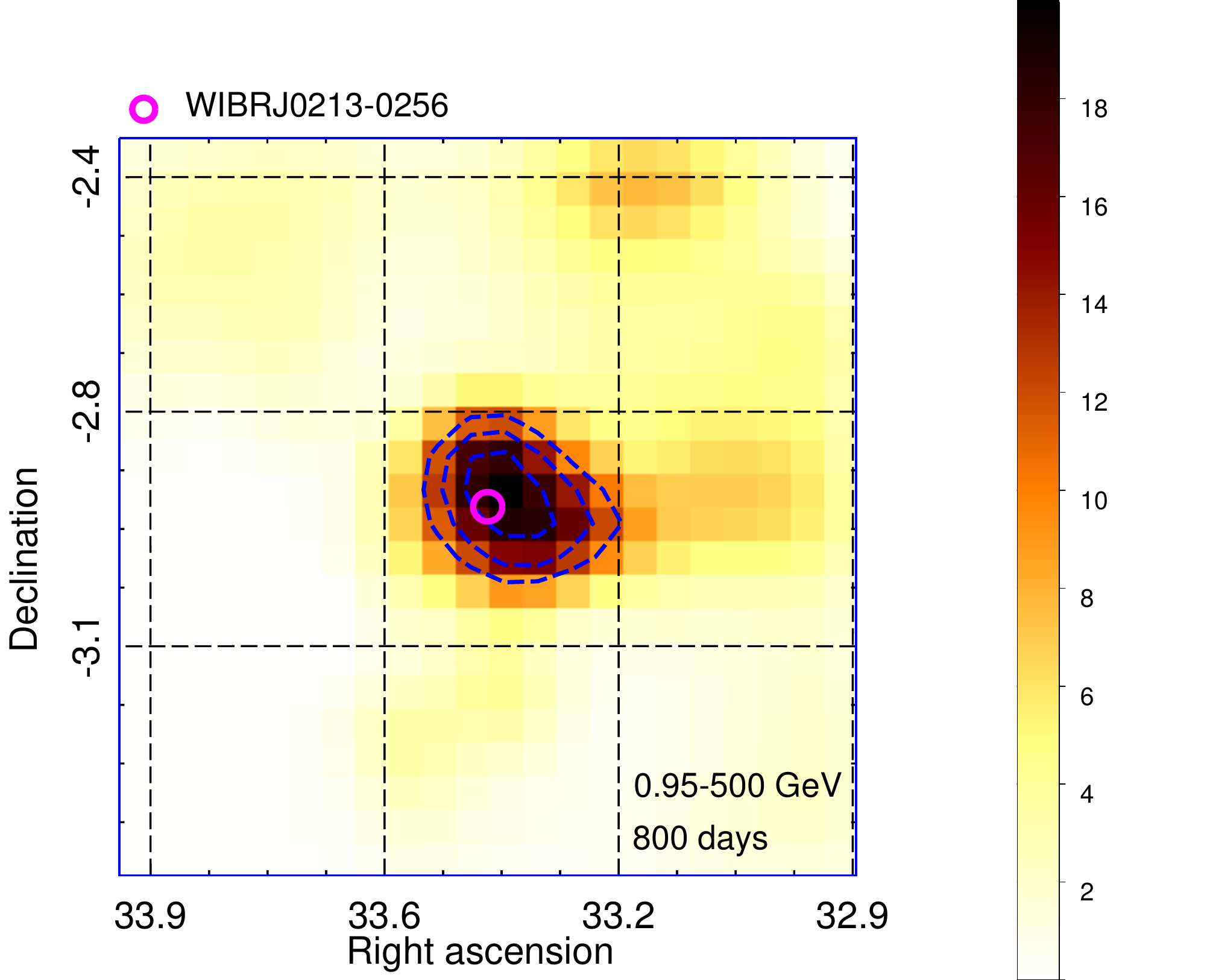} 
  \includegraphics[width=2.4in]{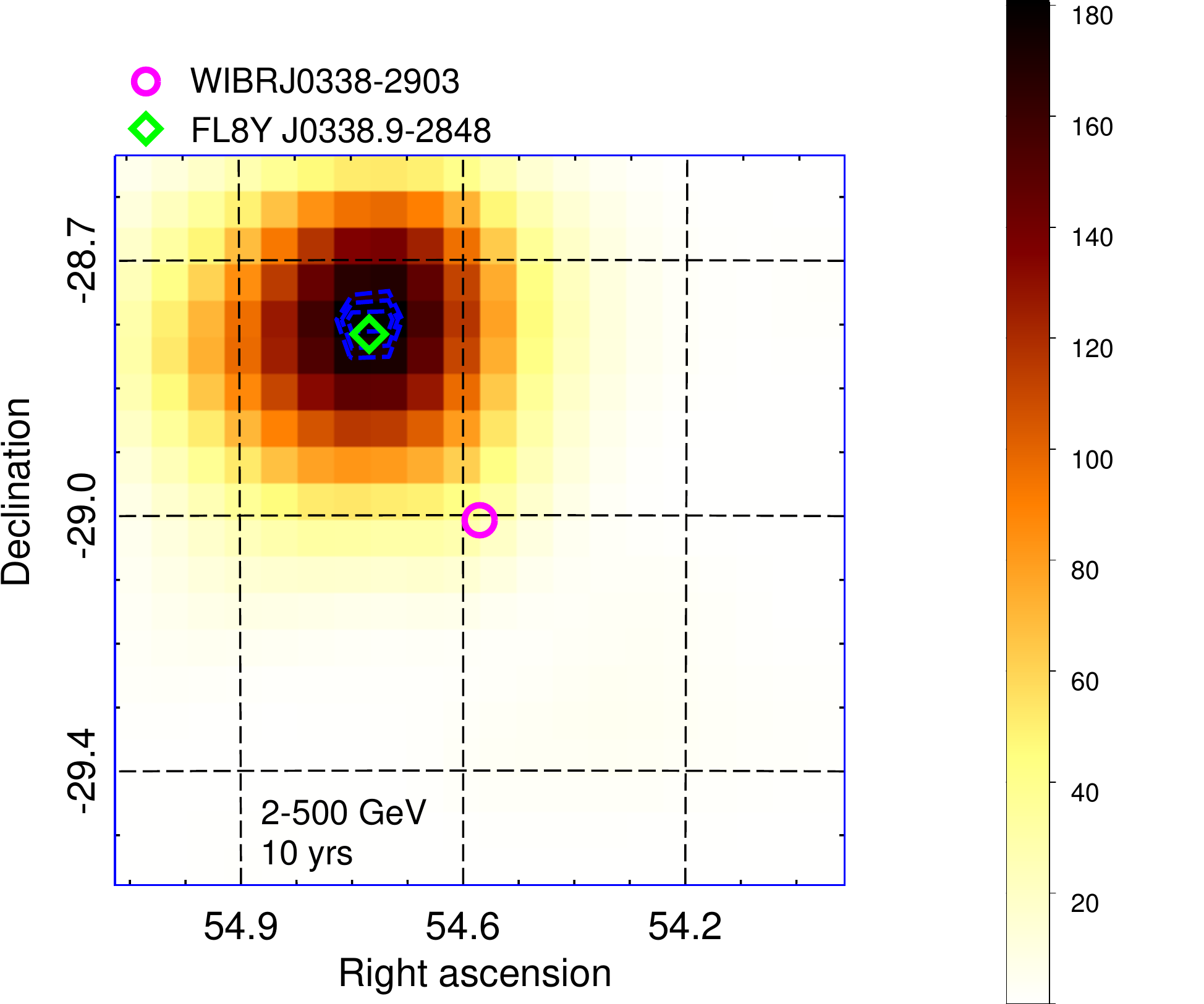}
  \caption{We show the TS map corresponding to a new detection of WIBRJ0213-0256 (left plot) and WIBRJ0338-2903 (right plot). The WIBR seed positions 
are shown by the magenta circles. In the right plot, the green diamond indicates the nearby Fermi source. The blue dashed contour lines correspond to 
68\%, 95\%, and 99\% containment region for the gamma-ray signature position (from inner to outer lines) in both TS maps.}
    \label{fig3}
 \end{center}
 \end{figure}

A total of 15 sources were found to be promising gamma-ray candidates. In some cases, the TS peak corresponded to closeby 4FGL sources or there was 
an FGL source within the region. In other cases, the signatures were identified as spurious. One of the promising gamma-ray candidates is shown in 
the right panel of Figure \ref{fig3}.
\vspace{3.3cm}


\begin{thebibliography}{}


\bibitem[Abdo  \etal\ (2010)]{abdo2010a}
{Abdo A. A., et al., } 2010, 
\textit{ApJS}, 188, 405

\bibitem[Abdo et al. (2010)]{abdo2010b}
{Abdo A. A., et al.,} 2010, 
\textit{ApJ}, 720, 435

\bibitem[Abdollahi et al. (2020)]{4fgl2020}
{Abdollahi S.,  et al.,} 2020,
\textit{ApJS}, 247, 33

\bibitem[Acero et al. (2015)]{acero2015}
{Acero F., et al.,} 2015, 
\textit{ApJS}, 218, 23

\bibitem[Ajello et al. (2015)]{ajello2015}
{Ajello M., et al.,} 2015, 
\textit{ApJ}, 800, 27

\bibitem[Arsioli et al. (2017)]{arsioli2017b1}
{Arsioli B. \& Chang Y.-L.,} 2017,
\textit{A\&A}, 598, 134

\bibitem[Arsioli et al. (2018)]{arsioli2018b2}
{ Arsioli B.  \& Polenta G., } 2018,
\textit{A\&A}, 616, 20

\bibitem[Chang  \etal\ (2017)]{chang2017}
{Chang Y.-L., Arsioli B., Giommi P. \& Padovani P., } 2017,
\textit{A\&A}, 598, 17

\bibitem[D'Abrusco et al. (2014)]{abrusco2014}
{D'Abrusco R., et al.,} 2014, 
\textit{ApJS}, 215, 14

\bibitem[Massaro et al. (2015)]{massaro2015}
{Massaro E., et al.,} 2015, 
\textit{Ap\&SS}, 357, 75

\bibitem[Mattox  et al. (1996)]{mattox1996}
{Mattox J. R., et al.,} 1996, 
\textit{ApJ}, 461, 396

\bibitem[Mukherjee et al. (1997)]{mukherjee1997}
{Mukherjee R., et al.,} 1997, 
\textit{ApJ}, 490, 116
 
\bibitem[Urry  \etal\ (1995)]{urryp1995}
{Urry C. M., \& Padovani P., } 1995,
\textit{PASP}, 107, 803

\end{thebibliography}
\end{document}